\documentclass[useAMS,usenatbib,usegraphicx]{mn2e}

\usepackage{times,xspace}

\def\lya{Ly$\alpha$\xspace}

\def\kms{\ensuremath{\mathrm{\,km\,s}^{-1}\xspace}}

\def\al2{Al{\sc ii}$\lambda$167.0}
\def\si2{Si{\sc ii}$\lambda$152.6}
\def\mg2{Mg{\sc ii}$\lambda\lambda$279.6,280.3}

\def\heii{He{\sc ii}\xspace}

\def\Hi{H{\sc i}\xspace}
\def\Hii{H{\sc ii}\xspace}
\def\hii{H{\sc ii}\xspace}

\def\hubble{\hbox{$h_{70}$}}
\def\h70m1{\hbox{$h_{70}^{-1}$}}
\def\kmsmpc{\hbox{km s$^{-1}$ Mpc$^{-1}$}}
\def\hm2{\hbox{$h_{70}^{-2}$}}

\def\msun{\ensuremath{M_{\odot}}}

\def\la{\mathrel{\mathchoice {\vcenter{\offinterlineskip\halign{\hfil
$\displaystyle##$\hfil\cr<\cr\sim\cr}}}
{\vcenter{\offinterlineskip\halign{\hfil$\textstyle##$\hfil\cr
<\cr\sim\cr}}}
{\vcenter{\offinterlineskip\halign{\hfil$\scriptstyle##$\hfil\cr
<\cr\sim\cr}}}
{\vcenter{\offinterlineskip\halign{\hfil$\scriptscriptstyle##$\hfil\cr
<\cr\sim\cr}}}}}
\def\ga{\mathrel{\mathchoice {\vcenter{\offinterlineskip\halign{\hfil
$\displaystyle##$\hfil\cr>\cr\sim\cr}}}
{\vcenter{\offinterlineskip\halign{\hfil$\textstyle##$\hfil\cr
>>\cr\sim\cr}}}
{\vcenter{\offinterlineskip\halign{\hfil$\scriptstyle##$\hfil\cr
>>\cr\sim\cr}}}
{\vcenter{\offinterlineskip\halign{\hfil$\scriptscriptstyle##$\hfil\cr
>>\cr\sim\cr}}}}}

\def\arcsec{\hbox{$^{\prime\prime}$}}
\def\utw{\smash{\rlap{\lower5pt\hbox{$\sim$}}}}
\def\udtw{\smash{\rlap{\lower6pt\hbox{$\approx$}}}}

\def\farcsec{\hbox{$.\!\!^{\prime\prime}$}}

\begin{document}

\title[The ISM and stellar populations of a LBG at z=3.8]{Medium-resolution spectroscopy 
of FORJ0332-3557: 
Probing the interstellar medium and stellar populations of a lensed 
Lyman-break galaxy at z=3.77\thanks{Based 
on observations made at the ESO VLT under programs 74.A-0536 and 78.A-0240}}
\author[R.A. Cabanac, D. Valls-Gabaud and C. Lidman]{
R\'emi A. Cabanac$^{1}$, David Valls-Gabaud$^{2}$ and Chris Lidman$^{3}$ 
\thanks{E-mail: remi.cabanac@ast.obs-mip.fr (RAC); david.valls-gabaud@obspm.fr (DVG); 
clidman@eso.org (CL)}\\
$^{1}$LATT, Universit\'e de Toulouse, CNRS, 57 Avenue d'Azereix, 65000 Tarbes, France\\
$^{2}$GEPI, CNRS UMR 8111, Observatoire de Paris, 5 Place Jules Janssen, 92195 Meudon Cedex, France\\
$^{3}$ESO, Vitacura, Alonso de Cordova, 3107, Casilla 19001, Santiago, Chile\\
}

\date{Accepted ... Received ... ; in original form ... }

\pagerange{\pageref{firstpage}--\pageref{lastpage}} \pubyear{2008}

\maketitle

\label{firstpage}

\begin{abstract}
We recently reported the discovery of FORJ0332-3557, a lensed Lyman-break galaxy  
at $z=3.77$ in a remarkable example of strong galaxy-galaxy gravitational lensing.
We present here a medium-resolution rest-frame UV spectrum of the source, which
appears to be similar to the well-known Lyman-break galaxy MS1512-cB58 at $z=2.73$. 
The spectral energy distribution is  consistent with a stellar population of less 
than 30\,Ma, with an extinction of $A_V=0.5$ mag and an extinction-corrected star 
formation rate $\mathrm{SFR}_{UV}$ of 200--300$\,\h70m1\, \msun\,a^{-1}$. The Lyman-$\alpha$ line
exhibits a damped profile in absorption produced by a column density of about  
 $N_{\mathrm{HI}} = (2.5\pm 1.0)\times10^{21}$\,cm$^{-2}$, superimposed on an emission
line shifted both spatially (0\farcsec5 with respect to the UV continuum source) 
and in velocity space ($+830 \kms$ with respect to the low-ionisation absorption lines
from its interstellar medium), a clear signature of outflows with an expansion velocity 
of about 270\,\kms. A strong emission line from \heii$\lambda 164.04$nm indicates the
presence of Wolf-Rayet stars and reinforces the interpretation of a very young
starburst. The metallic lines indicate sub-solar abundances of elements Si, Al, and C 
in the ionised gas phase.
\end{abstract}


\pagerange{\pageref{firstpage}--\pageref{lastpage}}
\pubyear{2008}

\section[]{Introduction}
Very few high-redshift galaxies ($z>3$) have been observed 
spectroscopically at medium and high resolution until now
because they are generally too faint for 8/10-m class telescopes 
equipped with medium resolution spectrographs.
Over the past years the most successful technique to probe these 
high-redshift galaxies consisted in building composite spectra 
of low signal-to-noise samples \citep{shapley01,shapley03,noll04} and 
redshift-selected sub-samples \citep{steidel01,ando04}. 
Very recently, high-redshift gamma-ray bursts
have also led to a wealth of low-resolution spectral  data 
on the interstellar medium of young star-forming 
regions in high-redshift galaxies \citep{berger06,fynbo06,vreesvijk04,vreesvijk06}.
From these composite samples a picture emerges of very young 
and massive stellar populations of 1/10th- to 1/4th- solar 
metallicities, dominated by on-going star formation, with strong outflows, and a
dusty component within abundant neutral gas.

As of today, only a handful of galaxies are bright enough 
to be studied spectroscopically at medium and high resolution, and 
among them the most well-known example is MS 1512-cB58 (cB58) 
\citep[$z=2.73$; look-back time=11.4\,Ga]{pettini00,pettini02,savaglio02}.
A more recent example is the 6$L_*$ $z=5.5$ starburst \citep{dowhygelund05} discovered 
in the field of a $z=1.24$ galaxy cluster. 
The lensed galaxy FORJ0332-3557 \citep[$z=3.773$, look-back time=12.1\,Ga]{cabanac05}, 
which is the object of the present study, is the most recent addition to this 
select few\footnote{The advent of 
systematic surveys on wide fields is yielding many more potential 
candidates, see e.g. \citet{bolton06,cabanac07}.}.  
Because the galaxy is magnified 13 times by a foreground lens at $z \sim 1$, 
it is bright enough to be observed at medium to high spectral resolution.  We present 
here a 9h-deep 
medium-resolution spectrum of the lensed galaxy taken with VLT/FORS2. 
The observations and data reduction are presented in Section 2 while Section 3 
outlines the properties of the emission and absorption lines and in Section 4
we analyse the spectral energy distribution and the underlying stellar populations. 

Throughout this paper we assume a flat FRW metric 
with $\Omega_\Lambda = 0.7$, $\Omega_M = 0.3$ and a Hubble constant normalised at   
\hubble $= 70$\,km\,s$^{-1}$\,Mpc$^{-1}$. Following the recommendations adopted by 
the IAU, the letter {\sl a} is used for the non-SI unit of year.

\section{Observations and data reduction}

\begin{figure}
\begin{center}
 \includegraphics[width=6cm]{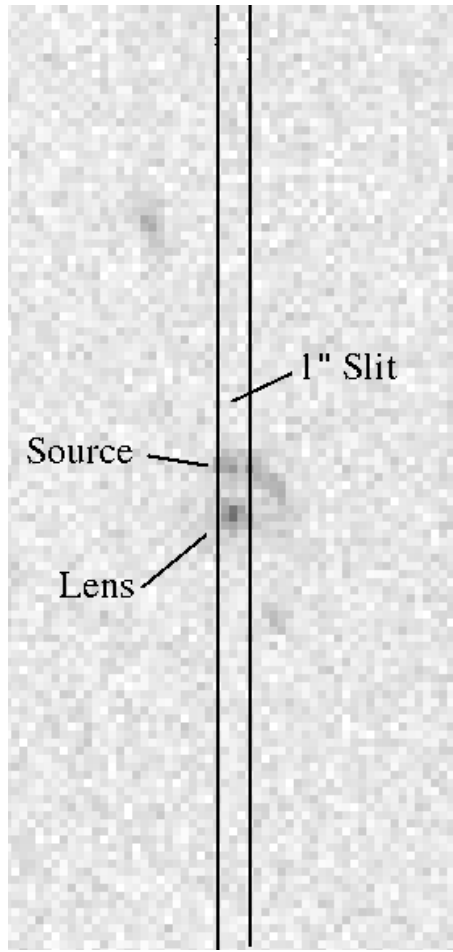}
\end{center}
 \caption{VLT/FORS2 acquisition frame (in the $R$ band) and 1\arcsec-slit mask 
position on FORJ0332-3557: the central object is the lens at $z\sim 1$ and 
the top arc is the lensed Lyman break galaxy at $z\sim 3.77$.\label{slit}}
\end{figure}

The observations were carried out with FORS2 on the European Southern 
Observatory VLT, under ESO programs 74.A-0536 and 78.A-0240. 
A total exposure time of 8.71 hours, split into 
6$\times$1495 s (1\arcsec slit) and 16$\times$1400 s 
(0\farcsec8 slit), was obtained (see Table~1). We used the holographic grism $600RI$ 
(0.163\,nm\,pixel$^{-1}$) together with the $GG435$ order sorting filter, which results in resolving powers
of 1200 and 1000 for the 0\farcsec8 the 1\arcsec\ slits, respectively. Since the
0\farcsec8 the 1\arcsec\ slits are slightly shifted with respect to each other in the focal plane, 
the resulting spectral coverage, 529.7-862.8\,nm  for the 1\arcsec slit and 501.1-832.5\,nm 
for the 0\farcsec8 slit, differ slightly. 
The detector was the upgraded MIT 2$\times$4096$\times$2048 mosaic, which with the standard 
resolution collimator yields a physical scale of 0\farcsec126\,pixel$^{-1}$. We oriented 
the slits North-South and centered them on the brightest part of the main arc (see Figure~\ref{slit}). 
We used the standard 2$\times$2 binned mode, which results in a final physical scale of 
0\farcsec252\,pixel$^{-1}$ and a sampling of 4 and 3.2 pixels for the 1\arcsec\ and 
0\farcsec8 slits, respectively. We placed the source on CCD1 (the upper northern chip), 
and used a 20\arcsec nod-on-slit strategy that allowed us to remove most sky artefacts.

The science data were taken during seven nights (2 nights for 1\arcsec\ slit, 5
nights for 0\farcsec8). The ESO standard calibration scheme delivers a set of bias, 
flatfield and wavelength calibration (from an He-Ar arc) frames for each night. 
We, then, subtracted master biases and divided by the 2-D normalised master 
flatfields. We evaluated the fringing contamination to be ca. 1\% 
peak-to-valley in the most sensitive part of the frames.

We performed two independent reductions on the 1\arcsec\ slit dataset. 
Firstly, we reduced separately each science exposure and recombined 
the resulting one-dimensional spectra after all calibrations. 
Secondly, we combined all science exposures 
using the FITS header World Coordinate info (WCS keywords) 
and removed the instrumental and sky artefacts on the combined 2-D image before
extracting and calibrating the source spectrum. 

In the first method we used standard NOAO/IRAF routines in order to 
extract the trace with a 4th-degree Legendre polynomial in the dispersion 
direction, weighted by the spectrum variance perpendicular to the dispersion 
direction. We removed the sky emission from 1-D modelling of the sky emission
as measured on both sides of the trace. We then calibrated each extracted 
trace with the wavelength calibration frames (1st order cubic spline) respecting 
the source spectrum slit position. FORS2 flexures are very small and the 
final rms of $\pm$0.02\,nm derived for the wavelength frame was assumed 
to apply to the science exposures as well. All wavelengths were shifted to the
heliocentric reference frame. Finally the combination of the resulting 6 
 1-D spectra (1\arcsec\ slit data; Table 1) was done using a median algorithm 
rejecting the minimum value and two maxima.

\begin{figure*}
 \includegraphics[width=16cm,angle=0]{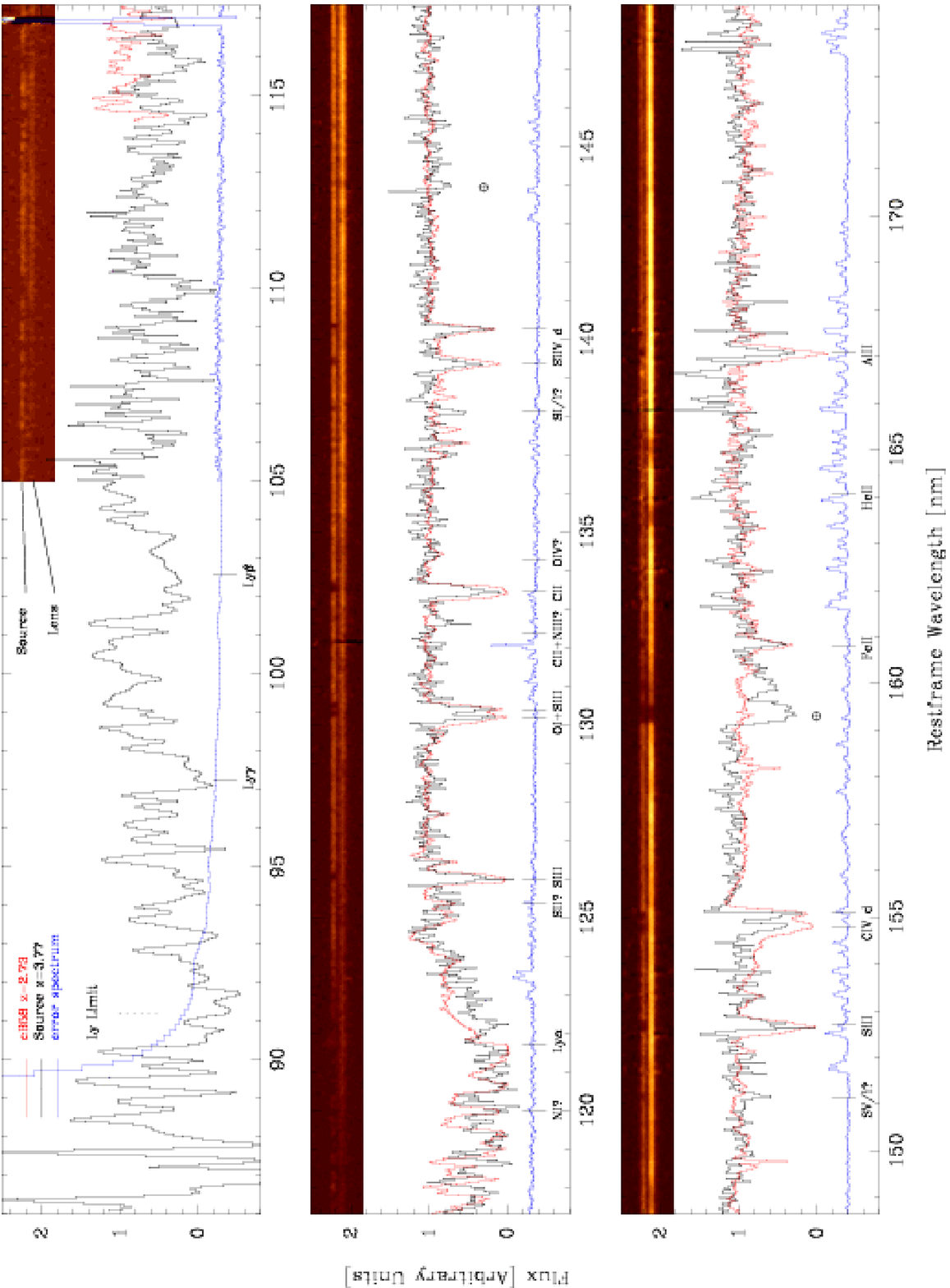}
 \caption{Continuum-normalised rest-frame ultraviolet spectrum of 
FORJ0332-3557 (black solid line), superimposed on  its error 
spectrum (blue solid line, offset for clarity) 
and the spectrum of the $z=2.73$ lensed galaxy cB58 (red solid line). 
The 2-D spectrum of the source, part of a compact lensing configuration, is 
also shown on the top.\label{spectrum}}
\end{figure*}

In the second method the sky substraction was done using an optimal fitting
while combining the 2-D spectra. The wavelength calibration frames proved 
to be stable between the two nights (1\arcsec\ slit data; Table~1). 
Figure~\ref{spectrum} shows the resulting 2-D frame after sky substraction.

Both methods lead to a similar final S/N and wavelength calibration 
accuracies across the useful wavelength range. The second method seems to lead 
to a slightly larger convolution of the observed line widths than the first
method. We reduced the 0\farcsec8 slit data using the second method only.

An important calibration step is the instrument response correction. 
The response curve of FORS2 was computed from spectrophotometric 
standard observations (LTT1788 and LTT2415 using a MOS 5\arcsec slit). 
Because the 0\farcsec8 and 1\arcsec slits are shifted with regards
to the MOS slit, the wavelength range of the standard covers a slightly different
range. We extrapolated the response curve linearly in the missing regions.

A final step was required to remove the contribution of the elliptical
galaxy that produces the gravitational lens effect on the 2-D spectrum of 
the source. We used two methods, which gave similar results. 
First, we subtracted the contribution of the lens measured 
on the southern profile at a symmetrical angular distance.
Second, we measured the dominant IR contribution of the lens over the arc 
and subtracted the best fit SED normalized to that contribution 
(cf. section \ref{pops}).

\begin{table}
 \centering
  \caption{Description of spectroscopic data with {\it VLT/FORS2}.}
  \begin{tabular}{@{}lccccc@{}}
  \hline
Date UT time & exp. time & airmass& seeing & slit\\
y-m-d h:m:s& s & - & \arcsec & \arcsec\\
 \hline
2004-11-16 01:02:19& 1495 & 1.446 & 0.56 & 1.0\\
2004-11-16 01:28:01& 1495 & 1.326 & 0.55 & 1.0\\
2004-11-16 01:54:29& 1495 & 1.231 & 0.65 & 1.0\\
2004-11-16 02:20:09& 1495 & 1.161 & 0.76 & 1.0\\
2005-01-30 01:20:45& 1495 & 1.103 & 0.50 & 1.0\\
2005-01-30 01:46:26& 1495 & 1.154 & 0.54 & 1.0\\
2006-10-16 06:02:29& 1400 & 1.028 & 0.66 & 0.8\\
2006-10-16 06:26:57& 1400 & 1.020 & 0.53 & 0.8\\
2006-10-16 06:51:33& 1400 & 1.021 & 0.61 & 0.8\\
2006-10-16 07:15:52& 1400 & 1.031 & 0.50 & 0.8\\
2006-10-16 07:55:52& 1400 & 1.068 & 0.73 & 0.8\\
2006-10-16 08:20:31& 1400 & 1.104 & 0.61 & 0.8\\
2006-10-21 07:37:59& 1400 & 1.070 & 0.48 & 0.8\\
2006-10-21 08:02:09& 1400 & 1.106 & 0.61 & 0.8\\
2006-11-21 07:03:26& 1400 & 1.266 & 0.85 & 0.8\\
2006-11-22 04:33:11& 1400 & 1.023 & 0.85 & 0.8\\
2006-11-22 04:57:20& 1400 & 1.036 & 0.89 & 0.8\\
2006-11-28 04:24:07& 1400 & 1.030 & 0.54 & 0.8\\
2006-11-28 04:48:36& 1400 & 1.048 & 0.63 & 0.8\\
2006-11-28 05:20:59& 1400 & 1.087 & 0.50 & 0.8\\
2006-11-28 05:45:14& 1400 & 1.130 & 0.57 & 0.8\\
2007-01-25 04:30:28& 1400 & 1.916 & 0.67 & 0.8\\
\hline
\end{tabular}
\end{table}

\section{Emission and absorption lines}
\subsection{Source redshift}
In the low-resolution discovery spectrum \citep{cabanac05} we estimated a source redshift of 
$z=3.773\pm0.003$ from a few strong interstellar lines. The improved signal-to-noise
obtained here (continuum peak S/N$\sim$15) allows us to 
measure here a revised systemic redshift of $z_{\mathrm{LIS}}=3.7723\pm0.0005$, 
using  11 stronger detections of 5 atomic species : C, O, Al, Si, and Fe, and 
excluding uncertain identifications and blended lines (see Table~2).
We marginally detect two important photospheric lines  O\,{\sc iv}\,$\lambda\,134.33$)
and S\,{\sc v}\,$\lambda\,150.18$ (see Fig.~\ref{spectrum}).
Hence we can only make a very tentative estimate of a possible velocity
offset between the  photospheric lines and the low-ionisation ISM lines of about 
$+110\pm30$\kms. This is about half the value inferred in cB58 
of $+210$\kms \citep{pettini00,pettini02}.

\subsection{The Lyman-$\alpha$ lines}
\label{lyman}
Figure~\ref{spectrum} shows the 2-D spectrum and the associated
1-D spectrum normalised to the continuum. Bluewards of
\lya the spectrum is poorly defined mostly because of foreground hydrogen absorption, 
and the  continuum has been assumed to be located at the peaks of the spectrum.
We also added the low-resolution VLT/FORS2 spectrum at wavelengths 
shorter than 110\,nm \citep{cabanac05}. 
The Ly$\alpha$, Ly$\beta$, and Ly$\gamma$ absorption features 
are conspicuous, and unresolved Ly$\delta$ absorption is detected
as well.

\begin{figure}
 \includegraphics[width=8.5cm,angle=0.0]{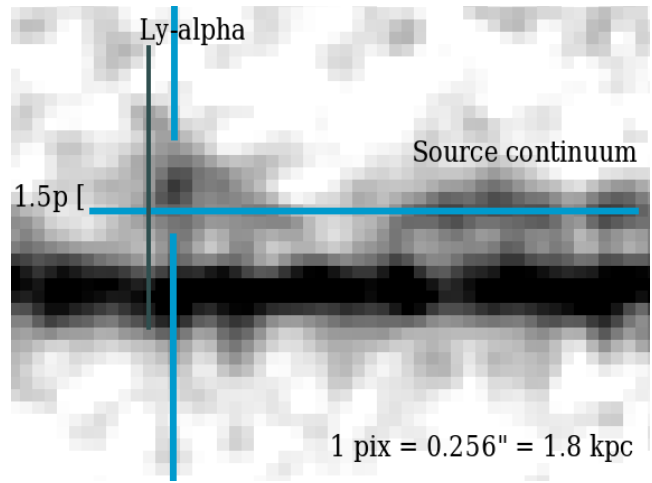}
 \caption{Zoom on the 2-D spectrum around \lya, showing the spatial
offset of $\sim 2\pm0.5$ pixels ($0.51\pm0.13$\,arcsec) and the
velocity shift of $\Delta v = +830$\kms, indicative of outflows (121.567 nm position
is the narrow green vertical line). 
 The lower trace is the spectrum of the elliptical galaxy that produces the
gravitational lensing effect.
\label{zoom}}
\end{figure}

Figure~\ref{zoom} zooms on the region around Ly$\alpha$. 
A faint emission peak at $\Delta v = +830$\,km/s of the Ly$\alpha$
absorption centroid and $\sim 2\pm0.5$ pixels ($0\farcsec51\pm0\farcsec13$) 
north of the continuum emission is clearly visible. This feature can
be interpreted either as a very bright \hii region in the periphery of the
main galaxy, or, more probably, as the signature of an expanding outflow. 
Assuming that the results of detailed 3-D Monte Carlo \lya radiative transfer 
codes \citet{verhamme06} (where the \lya~ emission is produced by an expanding
shell that is being ionised) can be applied to this case, we can infer
that the expansion velocity of the shell that gives rise to the \lya 
emission is constrained to be between $V_{exp}\sim \Delta v \sim $ 800 \kms 
(low column density case)
and $V_{exp} \sim \Delta v / 3 \sim$ 270 \kms (large column density case). It
is therefore essential to measure the \Hi~ column density in this context.

\begin{figure}
 \includegraphics[width=6.5cm,angle=-90]{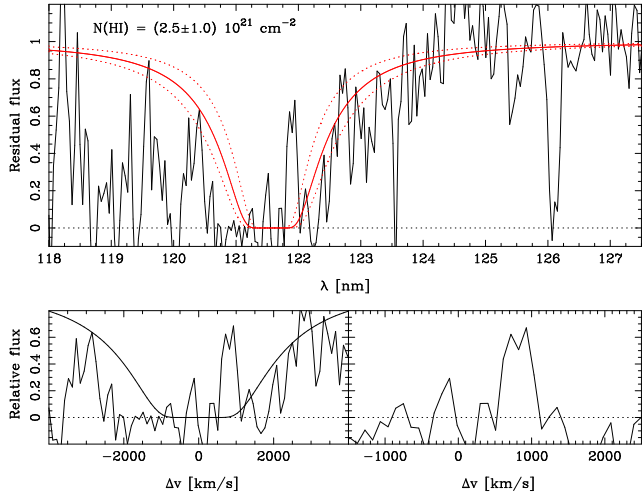}
 \caption{The best-fit Voigt profile corresponding to a column density 
of $(2.5\pm1)\times10^{21}$\,cm$^{-2}$ 
(solid and red dotted lines) is shown on the rest-frame spectrum (top frame).
The bottom frames show the same fit in velocity space $\Delta v$.
On the panel at the bottom right the model is subtracted out of the spectrum, 
showing a conspicuous emission feature peaking at $\sim 830$\,km/s. This feature
is also offset spatially by 2 pixels (ca. 0\farcsec5) from the main UV-continuum 
emitting source (see Fig.~\ref{zoom}). 
\label{lyman_alpha}}
\end{figure}

If we assume that the \lya absorption can be fitted by a damped Voigt
profile\footnote{This is clearly a rough approximation since the spectrum
comes from the integrated light of the source and there are photons that
are scattered into the line of sight. Hence the profile cannot be a pure
absorption profile, and the inferred column density is just indicative.}, we
 derive a column density of $N_{\mathrm{H{\sc I}}}=(2.5\pm1.0)\times10^{21}$
cm$^{-2}$. Figure~\ref{lyman_alpha} shows the best-fit damped Voigt profile and errors, 
along with the emission profile in velocity space obtained when the absorption
profile is subtracted. The column density is remarkably large, about 4 times
larger than in MS1215-cB58 \citep{pettini02,savaglio02}. If we adopt the model
of an expanding shell at constant velocity \citep{verhamme06}, this
column density still implies a large expansion velocity of about $V_{exp} \sim 270 $
km/s, similar to the one found in cB58 \citep{schaerer08}, and at the
upper end  of the Lyman-break galaxies \citep{shapley03} if they do have
large \Hi~ column densities. It is remarkable that in spite of the very
large column densities, such strong Lyman-$\alpha$ emissions are detected,
confirming that the interstellar medium is very clumpy, yielding smaller
effective optical depths which are very similar to the ones inferred in 
the low-redshift starbursts \citep{vallsgabaud93}, and allowing many \lya photons 
to escape.

The neutral gas content derived with Voigt profile fitting is fully consistent 
with the neutral gas content derived independently from the analysis of 
stellar populations (Section~\ref{pops}) with
 the gas-to-dust ratio taking Milky Way or LMC values, 
but is $\sim$ 5--10 times smaller than the values observed in the SMC.

\subsection{The Helium\,{\sc ii} $\lambda$164.04nm emission line: the signature of 
very massive stars}
\label{heii}
A broad emission feature at 164.04\,nm is seen in Fig.~\ref{spectrum}
at a redshift of $z=3.7738$ consistent with other faint photospheric features.
We associate this feature with the well-known 
 He\,{\sc ii} emission produced by Wolf-Rayet stars. Although not present in cB58, 
this feature  has already been detected in absorption in young LBGs 
\citep[][; cf. Fig.~15 and 16]{shapley01}, and in emission in the 
composite LGB spectrum of \citet{shapley03} but the origin
of the feature cannot be asserted by the authors because the composite 
spectrum mixes a wide range of stellar populations and ages. If 
FORJ0332-3557 is similar to other LBGs, it belongs to the strongly absorbed \lya quartile, 
and the case for a very young stellar population ($\sim$5\,Ma) of massive 
stars is strong. The He\,{\sc ii} feature of Fig.~\ref{spectrum} shows the typical asymmetric
 profile produced by strong 
stellar outflows of the evolved descendants of O stars more 
massive than $M > 20$–-$30 M_\odot$. The equivalent width
$W_0(\hbox{He\,{\sc ii}})$ yields the ratio of WR to O stars following \citet{schaerer98},
\begin{equation}
\log\left[\frac{\hbox{WR}}{\hbox{WR+O}}\right] = (-2.83\pm0.03) + 
(1.52\pm0.05)\log[W_0(\hbox{He{\sc ii}})]\; ,
\end{equation}
which for $W_0(\hbox{He\,{\sc ii}})=0.22\pm0.02$nm~ implies
a ratio O/WR$=5\pm1$.
In turn, this value sets a strong age limit on the stellar population of
$\la$6\,Ma rather independently of metallicity \citep{cervino94,schaerer03}. 
Stochastic effects due to the sampling of the 
stellar mass function (e.g. \citet{mcsdvg2003,mcsetal03}) imply that only the most
recent burst can be dated through this short-lived feature.

\subsection{Interstellar absorption lines}
The spectrum of FORJ0332-3557  shows characteristic absorption lines from 
starburst galaxies. They are summarised in Table~2. The 
medium-resolution spectrum of the LBG cB58 \cite{pettini00} is shown 
in Fig.~\ref{spectrum} in red, superimposed on our source, for a direct comparison 
of the insterstellar features. It is immediately obvious that the FORJ0332-3557 source 
is qualitatively similar to cB58. 

All common interstellar absorptions are found in FORJ0332-3557.
The strong absorption features include low-ionisation lines 
associated with neutral gas (Si\,{\sc ii} 
$\lambda\lambda\lambda$126.04 130.47 152.67, 
C\,{\sc ii} $\lambda$133.45, O\,{\sc i} $\lambda$130.22,
Al\,{\sc ii} $\lambda$167.08, Fe\,{\sc ii} $\lambda$160.84)
and high-ionisation lines associated with a hot gas phase 
(Si\,{\sc iv} $\lambda\lambda$139.38 140.28, 
C\,{\sc iv}$\lambda\lambda$154.82 155.08). Table~2 lists the ion line 
identification, 
vacuum rest-frame wavelength $\lambda_{vac}^{lab}$ , 
observed wavelength $\lambda_{vac}^{obs}$, redshift $z$,  rest-frame 
equivalent width $W_0$, oscillator strengths $f$, column
density, ion abundance with respect to solar
[$X/H$]$_\odot$ and comments. 
Additional uncertain identifications are question-marked, the lines 
noted {\it i?} might belong to interlopers at unknown redshift(s). We 
emphasize that the derived $W_0$ are very sensitive to both sky 
subtraction and continuum normalisation, hence the systematic errors caused by the 
continuum normalisation have tentatively been computed and are shown to be close  
to photon counting errors, while the sky subtraction error is much more difficult to 
quantify. As a sanity check, we computed the equivalent widths of absorption lines in the spectrum
of cB58 and found our measurements to be fully consistent with those published 
by \citet{pettini02}.

No nebular emission lines are detected in the present spectrum other than
 He\,{\sc ii} $\lambda$164.04nm (\S~\ref{heii}). 
A weak detection of C\,{\sc iii}] $\lambda$ 190.87 seen on a previous
spectrum \citep{cabanac05} suggests that contamination
of the high-ionisation lines by nebular emission is
present but small. P-Cygni profiles are visible on 
C\,{\sc ii} $\lambda$133.45, and C\,{\sc iv}$\lambda$155.08.

There are several ways to derive the abundances in the interstellar
medium of distant galaxies 
\citep{spitzer78,pettini02,savaglio02}. Ideally one should 
build a curve of growth by fitting Voigt profiles and Doppler parameters $b$ 
for all ions independently. Because the resolution of our observed spectrum is just under the 
resolution one needs for Voigt profile fitting, 
and is penalised by a low signal-to-noise ratio, most of the strong lines appear
saturated, and most weak lines are dominated by noise. 

A careful analysis of the ISM metallicity goes beyond the present
paper and will be done elsewhere. Here we present only qualitative arguments
on the curve of growth, and Doppler parameters $b$. 
Assuming that the interstellar medium in FORJ0332-3557 is optically thin, 
one can infer lower limits to  
column densities, log($N$[cm$^{-2}$]), and abundances (given in  
Table~2) by taking the optically thin approximation  
\begin{equation}
\log N [\mathrm{cm}^{-2}] = 19.053+\log\left[\frac{W_\lambda}{\lambda^2 f}\right] \quad ,
\end{equation}
where $f$ is the line oscillator strength. The equivalent width, $W_\lambda$, 
and the wavelength, $\lambda$, are in nm.
A tentative curve of growth indicates that
the ion abundances could be 2-3 dex 
larger for a Doppler parameter of $b=50$\kms. In this context, the most
constraining line, besides Si{\sc ii}$^*\lambda$153.3 which may be blended,
is Fe{\sc ii}$\lambda$160.8, which appears unsaturated and 
whose small equivalent width is similar to the one measured in cB58 and would
yield $b \sim 60$\kms~, similar to the $b \sim 70$\kms reported in cB58 \citep{pettini02}.

Compared to cB58, FORJ0332-3557 $W_0$ are lower by factors of 2-3 
(C\,{\sc iv}$\lambda\lambda$155.08 155.08, 
Al\,{\sc ii}$\lambda$167.08, O\,{\sc i}$\lambda$130.22) to a 
factor of 1-1.2 (Si\,{\sc iv}$\lambda\lambda$139.38 140.28, 
Fe\,{\sc ii}$\lambda$160.84).
More detailed analyses on elemental abundances and depletion in 
the interstellar medium of FORJ0332-3557 will be presented elsewhere.

\begin{table*}
\label{table:abundances}
 \centering
 \begin{minipage}{180mm}
  \caption{Interstellar absorption lines}
  \begin{tabular}{@{}lcccccccl@{}}
  \hline
Ion & $\lambda_{vac}^{lab}$ & $\lambda_{vac}^{obs}$ & redshift & 
$W_0$\footnote{Errors are given as statistical ($\pm$photon noise) and systematic ($^+_-$systematic
 uncertainty in location of the continuum)} & 
$f$ & log$N$\footnote{Lower limits based on the assumption of an optically thin medium}&
[$X/H$]$_\odot$\footnote{Assuming solar abundances from \citet{asplund05}}&Comments\\
    &       nm               &            nm        &          &                                        nm                                            & &cm$^{-2}$& &\\
 \hline
H\,{\sc i} Ly$\gamma$& 97.254 & - & - & $1.50\pm0.30^{+0.20}_{-0.10}$  &  0.0290 &-&-&\\
H\,{\sc i} Ly$\beta$& 102.57 & - & - & $2.10\pm0.80^{+0.80}_{-0.60}$  & 0.07912 &-&-&\\
H\,{\sc i} Ly$\alpha$& 1215.7 & - & - & - & 0.4164 &$21.4\pm0.2$ &-&\\
He\,{\sc ii} & 164.04 & 783.09 & 3.7738 & $-0.22\pm0.01^{+0.02}_{-0.02}$ & - & - & - & WR feature\\
C\,{\sc ii} & 133.45 & 637.01 & \it 3.7732& $0.21\pm0.03^{+0.03}_{-0.02}$ & 0.1278  &$15.019\pm0.10$ &$>-2.8$&blended with C\,{\sc ii*}$\lambda$133.57 \\
C\,{\sc iv} & 154.82 & 738.74 & 3.7716& $0.10\pm0.03^{+0.06}_{-0.03}$&  0.1908 &$14.40\pm0.21$&$>-3.6$&\\
C\,{\sc iv} & 155.08 & 740.00 & \it 3.7718& $0.19\pm0.05^{+0.02}_{-0.02}$& 0.09522 &$14.97\pm0.25$&$>-3.0$& sky contamination\\
N\,{\sc i}? & $\sim 120$ & 572.70 & \it3.7718& $0.36\pm0.18^{+0.08}_{-0.05}$& 0.04023 &$15.84\pm0.25$&$-1.6?$&triplet 
N\,{\sc i}$\lambda$ 119.95 120.02 120.07\\
N\,{\sc iii}? & 132.43& 632.24 & \it 3.7741 & - & - & - & - & blended with C\,{\sc ii}$\lambda$132.39\\
O\,{\sc i} & 130.22 & 621.31 & \it3.7714 & $0.15\pm0.03^{+0.05}_{-0.03}$& 0.04887 &$15.30\pm0.17$&$>-3.0$&blended with Si\,{\sc ii}$\lambda$130.44\\
O\,{\sc iv} & 134.34 & 641.27 & 3.7735 & $0.14\pm0.01^{+0.05}_{-0.03}$&-&-&-& photospheric\\
Al\,{\sc ii}& 167.08 & 797.41 & \it3.7726 & $0.19\pm0.02^{+0.07}_{-0.03}$&1.833 &$13.62\pm0.22$&$>-2.5$&sky contamination\\
S\,{\sc i}/i? & 138.16 & 659.29 & 3.7719& $0.11\pm0.01^{+0.03}_{-0.02}$&-&-&-&\\
S\,{\sc ii}/i? & 125.38 & 598.57 & 3.7740& $0.12\pm0.02^{+0.04}_{-0.05}$& 0.01088 &$15.90\pm0.15$&$-1.1?$&\\
S\,{\sc ii}/i? & 125.95 & 601.31 & \it3.7742& $0.19\pm0.04^{+0.03}_{-0.02}$& 0.01624 &-&-&blended with Si\,{\sc ii} $\lambda$126.04 \\
S\,{\sc v}? & 150.18 & 716.39 & \it3.7741&$0.3\pm0.03^{+0.04}_{-0.2}$&0.00545&-&-& photospheric\\
S\,{\sc ii}/i? & 151.12 & 721.43 & \it3.7740& $0.07\pm0.01^{+0.02}_{-0.02}$&-&-&- &blended with Si\,{\sc ii}$\lambda$151.21 \\
Si\,{\sc ii} & 126.04 & 601.84 & \it3.7749& $0.19\pm0.04^{+0.03}_{-0.02}$& 1.007 &$14.13\pm0.11$&$>-2.8$&blended with S\,{\sc ii}$\lambda$125.95 \\
Si\,{\sc ii} & 130.47 & 622.49 & \it3.7710& $0.16\pm0.01^{+0.06}_{-0.03}$& 0.094 &$15.05\pm0.37$& $-1.9?$&blended with O\,{\sc i}$\lambda$130.22\\
Si\,{\sc ii} & 152.67 & 728.39 & 3.7710& $0.24\pm0.03^{+0.02}_{-0.02}$&  0.130 &$14.96\pm0.11$&$>-2.0$&\\
Si\,{\sc ii*}?& 153.34 & 731.40 & 3.7697& $0.04\pm0.01^{+0.01}_{-0.01}$ &0.132&$14.20\pm0.10$&$-2.8?$&\\
Si\,{\sc iv} & 139.38 & 665.09 & 3.7718& $0.19\pm0.03^{+0.03}_{-0.02}$&  0.5140 &$14.33\pm0.10$&$>-2.6$&\\
Si\,{\sc iv} & 140.28 & 669.38 & \it3.7718&$0.21\pm0.03^{+0.04}_{-0.02}$& 0.2553&$14.66\pm0.12$& $>-2.3$&blended with S\,{\sc i}$\lambda$140.15?\\
Fe\,{\sc ii} & 160.84 & 767.68 & 3.7728& $0.12\pm0.02^{+0.01}_{-0.02}$& 
0.058 &$14.95\pm0.11$&$-2.0$&\\
\hline
\end{tabular}
\end{minipage}
\end{table*}

\section{Stellar populations}\label{pops}
After analysing the absorption and emission lines here we study the properties
of the continuum, using recent stellar synthesis codes, which, for the first
time, predict  high-resolution spectra of starburst populations. We used both {\sc{starburst99}}
\citep{leitherer99,vazquez05} and {\sc{sed@}}
\citep{buzzoni89,cervino02,cervino04} in an attempt 
to derive plausible combinations of ages, metallicities and extinctions 
by dust from the FORJ0332-3557 ultraviolet spectrum (Fig.~\ref{spectrum}). 

\subsection{Spectra predicted by \sc{starburst99}}
\begin{figure*}
 \includegraphics[width=\textwidth]{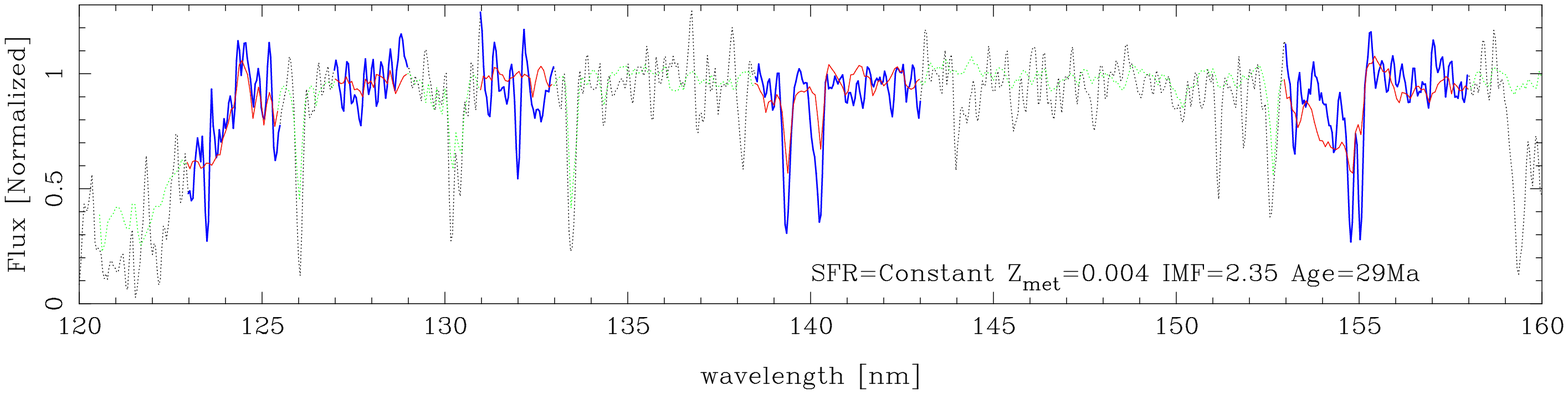}
\\[3mm]
\includegraphics[width=\textwidth]{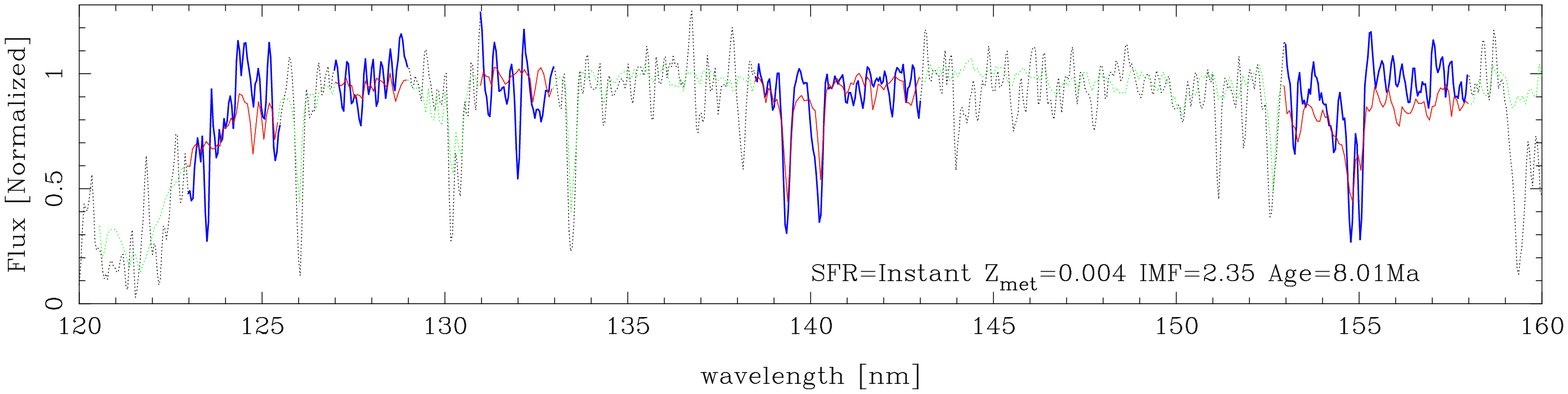}
 \caption{The best fit {\sc{starburst99}} models (red lines) are overlaid on the normalised 
spectrum of FORJ0332-3557 source (blue line) for two extreme scenarios of star formations
with a Salpeter IMF from 1-100\,M$_\odot$, at a metallicity $Z$=0.004 (1/5
$Z_\odot$). The parts of the spectrum excluded of the fit are shown as dotted
lines (see text). The top frame shows a constant SFR for 29\,Ma. 
Ages older than 20\,Ma are strongly favoured. The bottom frame shows an 8\,Ma-old 
instantaneous burst of star formation. No model (at any available 
metallicities (1/20--1\,$Z_\odot$) can reproduce the observed depths of the  
lines. For a constant SFR, larger metallicities produce older ages 
(29\,Ma at $Z$ = 0.001, $>$49\,Ma at $Z$ = 0.008). Instantaneous burst models 
tend to produce deeper lines while  constant SFR models tend to better fit the
observed P-Cygni profiles of C\,{\sc iv}$\lambda$155.08 nm.\label{sb99}}
\end{figure*}

We build two series of {\sc{starburst99}} spectra using (1) an IMF with a Salpeter slope from
1-100 M$_\odot$\footnote{Tests where the upper stellar mass cutoff of the IMF is
is reduced to 30\,M$_\odot$ 
give very similar best-fit ages.}, (2) a self-consistent association of low 
metallicity tracks with LMC/SMC atmospheric stellar templates, and (3) solar 
metallicity tracks with galactic atmospheric stellar templates, from ages 0.01\,Ma 
to 50\,Ma, for metallicities $Z$ of 0.001 (1/20$Z_\odot$), 0.004 (1/5$Z_\odot$), 
0.008 (2/5$Z_\odot$), and 0.020 ($Z_\odot$). 
All {\sc{starburst99}} models were fitted to the spectrum of Fig.~\ref{spectrum},
pre-filtered with a Savitsky-Golay filter (conserving the equivalent-width, 
FWHM and line profiles), to the same resolution as the {\sc{starburst99}} 
synthetic templates. The region below 123\,nm was excluded from the fits, 
because the source is heavily contaminated by foreground interlopers.
For solar metallicities, {\sc{starburst99}} high-resolution atmospheric 
templates go up to 180\,nm, but only 160\,nm for subsolar SMC/LMC 
metallicities. Only the highly ionised absorption lines were
considered for the fitting, to avoid excessive interstellar contamination.
An additional range of wavelengths (near 137 and 144\,nm) was excluded 
due to telluric contamination. 
The regions included in the fit are shown as solid lines in Fig.~\ref{sb99}.
Different sets of constraints can be derived by fitting the line 
equivalent widths of the normalized spectrum and by fitting the shape 
of the spectrum continuum. Fig.~\ref{sb99} compares the best
fit models with the normalised spectrum for two extreme scenarios of star 
formation: (i) a single burst of star formation (bottom) and (ii) a continuous 
SFR (top). The best fit models for $Z$=0.004 have ages of 8\,Ma (single
burst) and 29\,Ma (constant SFR). The age of the best fit models is found to
be insensitive to metallicity in the case of single bursts. 
In the case of a constant SFR, increasing the metallicity tends to increase the age of the 
best fit model :  29\,Ma at $Z$=0.001, $>49$\,Ma at $Z$=0.008, and more 
than 50\,Ma at $Z$=0.02. We explored solar metallicity models with 
ages of 200\,Ma and although the reduced $\chi^2$ decreased slightly
the highly-ionised absorption lines did not get deeper significantly.
The reduced $\chi^2$ shows that constant SFR models become insensitive 
to age above 20\,Ma, whereas single burst models appear to isolate a well-defined small region 
(6-9\,Ma). However, neither of these scenarios 
can reproduce the observed depth of the absorption lines, which probably
argues for interstellar origin for the highly ionised lines.
Finally, the single burst scenario seems to produce slightly deeper 
absorption lines, whereas the constant SFR scenario better fits the 
P-Cygni profiles of C\,{\sc iv}$\lambda\lambda$155.08\,155.08\,nm 
and the red side of \lya. We also tested different stellar wind models 
without any significant differences.

\subsection{Spectra predicted by \sc{sed@}}
\begin{figure*}
 \includegraphics[height=4.7cm]{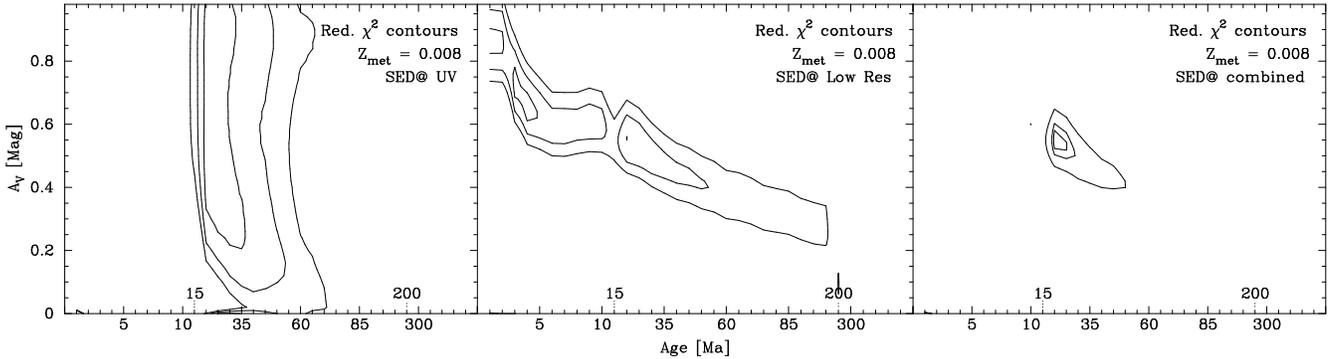}
 \caption{Reduced $\chi^2$ contours are given for the best fit
$Z$=0.008 {\sc{sed@}} models for a grid of ages and extinctions.
The contour levels are arbitrarily chosen to be 1, 3 and 10\% of the 
reduced $\chi^2$ minimum, to outline the trends.
The left panel shows the contours derived for a continuum-normalised
set of deep absorption lines alone (cf. text), the center panel shows the 
constraints derived from the photometric colours alone, and the right
panel shows the combined contours of the two independent sets taken together.
The optimal parameter set is age=20$\pm5$\,Ma, $A_V=0.55\pm0.02$\label{seda}.
}
\end{figure*}

The stellar population synthesis code {\sc{sed@}} is included in the {\it Legacy Tool  project} 
of the {\it Violent Star Formation Network}\footnote{See the {\sc{sed@}} online  
Reference Manual at {\tt http://www.iaa.es/\~{}mcs/sed@/index.html} 
for more information.} \citep{buzzoni89,cervino02,cervino04}. It computes 
UV spectra with the highest resolution currently available, consistently 
computing the errors introduced by a limited number of bright stars in a 
small burst (this effect can probably be neglected in our case unless the
lens amplifies a single star-forming region within the source).
We use a series of instantaneous burst models from 0 to 10\,Ga, a Salpeter IMF, and  
Geneva stellar tracks at 4 metallicities:
$Z$ of 0.001 (1/20$Z_\odot$), 0.004 (1/5$Z_\odot$), 0.008 (2/5$Z_\odot$), 
and 0.020 ($Z_\odot$). As in the previous section, we treat independently
the constraints coming from the normalised UV absorption features,
and the overall photometric SED of the source.
For the UV-spectrum fitting, we isolate the strongly-ionised absorption 
features containing most of the signal and exclude the weakly-ionised
features dominated by interstellar features, using the wavelength intervals 
120--125, 127--129, 131--133, 134--143, 145--152, and 153--158\,nm. We normalise 
each interval to the continuum by dividing each wavelength bin by the 
linear interpolation between the two extreme values of the interval. 
We compute the reduced $\chi^2$ between the templates and the source UV 
spectrum for a  grid of extinctions ($0<A_V<1$), and ages (1-1000\,Ma).
The photometric SED of the templates are integrated over the instrumental
transmission curves (filter+telescope+CCD detector) given by VLT/FORS1 ($B$, 
$V$ and $R_C$ bands) and VLT/ISAAC ($J$, $H$, $K_S$ bands) manuals, normalised
to VEGA magnitude using  Kurucz's solar model \citep{castelli94}.
Figure~\ref{seda} shows the resulting reduced $\chi^2$ contours. Ages are along
the horizontal axis (steps of 1\,Ma from 0 to 10\,Ma, steps of 5\,Ma from 10 to 100, 
steps of 100\,Ma from 100\,Ma and older ages). The extinction $A_V$ is given on the
vertical axis. The left panel shows the contours of $\chi^2$ values obtained
in the fitting of the UV features described above 
(the contour levels are given at 1, 3 and 5\% above the value of the minimum $\chi^2$). 
The central panel shows the contours obtained in the fitting of the photometric SED (contour levels
at 1.5, 3, and 5 times the minimum $\chi^2$), while the right panel shows 
the combined constraints from both the overall SED and the UV features 
(contour levels at 1\, 3, and 10\% above the minimum $\chi^2$). 
The left frame shows a complete degeneracy in extinction, as expected since we
normalise the continuum and the UV is no longer sensitive to extinction. 
The central panel shows the well-known degeneracy
between age and extinction but clearly shows a tendency to prefer young
ages with higher extinction, rather than older ages with small extinctions.
It should be noted that the $\chi^2$ of the UV fitting alone shows a minimum at  
$Z$=0.004, whereas the $\chi^2$ of SED-fitting alone favours the smallest
metallicity models $Z$=0.001. The combined set leads to $Z$=0.008 
(Fig.~\ref{seda}) but it is clearly not a very constraining result due the degeneracies
involved\footnote{See for example \citet{schaererpello2005} for other examples
of degeneracies in the SED fitting of high-$z$ galaxies.}. Because the two sets of 
constraints are independent and almost orthogonal, the combined $\chi^2$ shown in
the right panel of Fig.~\ref{seda} (each $\chi^2$ is given the same weight) 
gives a tighter constraint on the age-$A_V$ parameter space. The combined 
best fit parameters ($Z$, age [Ma], $A_V$[mag])  are (0.001, 15, 0.62), (0.004, 20, 0.54),
(0.008, 20, 0.54), (0.020, 10, 0.64), (0.040, 15, 0.52) for reduced $\chi^2$
values of 0.853, 0.800, 0.788, 0.792, 0.789.

In conclusion, both stellar synthesis codes give a consistent picture
of the global properties of the source galaxy in FORJ0332-3557, although they marginally
disagree on details. They both find  young stellar populations of 
$<30$\,Ma and the presence of some dust extinction. Neither the metallicity, 
nor the IMF can be robustly constrained although {\sc{sed@}} marginally
favours $Z$=0.008. We emphasize that none of the models, 
at any metallicity and age, can fit the depth of the observed 
strongly-ionized absorption lines.
{\sc{starburst99}} points to a young stellar population of 8--10\,Ma,
in the case of an instantaneous burst, and a few tens of Ma, in case of a constant SFR,
whereas {\sc{sed@}} models seem to favour slightly older ages (10--20\,Ma) for an 
instantaneous burst. Lastly, {\sc{starburst99}} tends to give a better
fit of the absorption features than {\sc{sed@}}, including interstellar features
of course because {\sc{sed@}} uses synthetic stellar spectra whereas 
{\sc{starburst99}} uses observed stellar templates where the Milky Way
interstellar absorption features have not been removed. 
Finally, {\sc{sed@}} over predicts an absorption line 
Si\,{\sc ii}$\lambda$ 126.5\,nm and a P-Cygni profile of 
O\,{\sc i}$\lambda$130.22\,nm.
Finally, the ages derived by stellar synthesis are older than the
presence of the Helium\,{\sc ii} emission line seems to suggest.
This is not contradictory but rather suggests that the two analyses
are not sensitive to the same features, which probably come from different
regions of the galaxy and may point at a range of ages and multiple
star formation episodes.

\subsection{Integrated spectral energy distribution and dust extinction}
\begin{figure}
 \includegraphics[width=5.5cm,angle=-90]{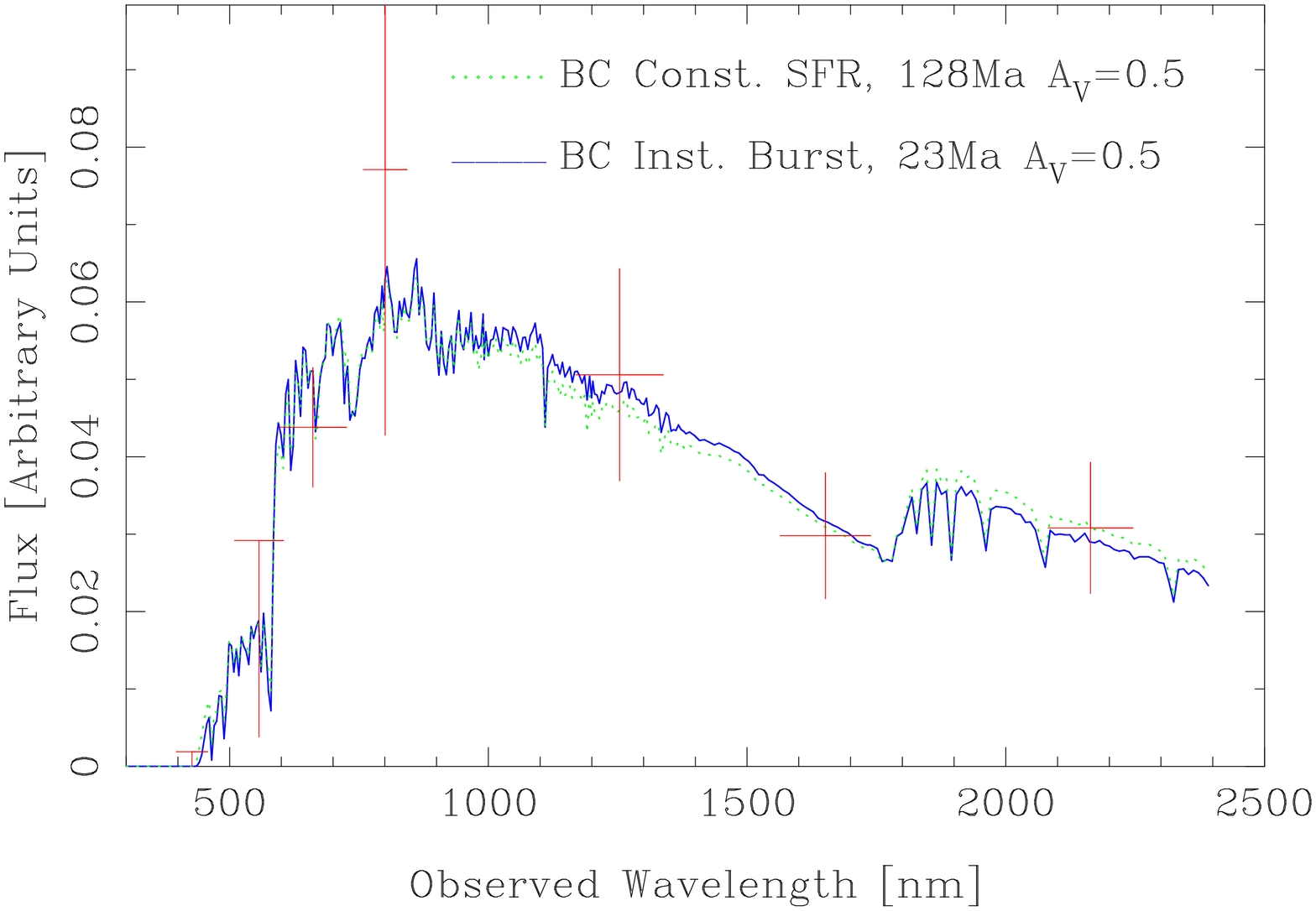}
\\[4mm]
\includegraphics[width=5.5cm,angle=-90]{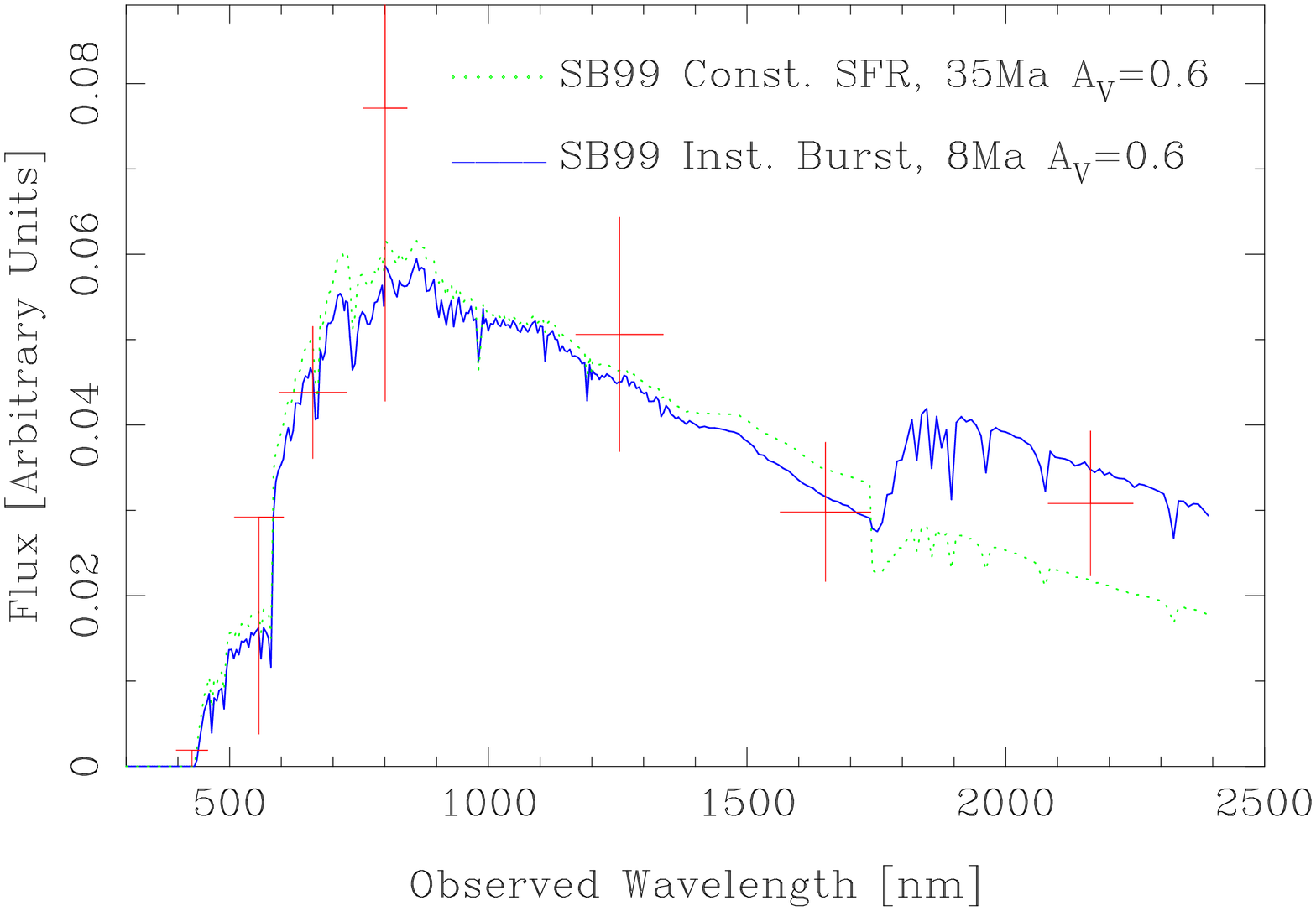}
 \caption{The $BVRIJHK_s$ SED of FORJ0332-3557  (red crosses, the 
vertical bars indicating systematic errors and the horizontal ones giving 
the spectral bandwith of each filter). 
The top frame shows {\sc{HyperZ}} best fit  solar metallicity
templates: single-burst  template of age 23\,Ma (blue solid line) and 
constant SFR followed by passive evolution  template of age 
128\,Ma (green dotted line). Both templates are the best fit with an extinction  
$A_V=0.5$\,mag. The bottom frame shows {\sc{HyperZ}} best fit with {\sc{starburst99}} 
$Z$=0.004 templates:  instantaneous burst  
of age 8\,Ma (blue solid line) and constant SFR  of age 35\,Ma 
(green dotted line), both with a best fit extinction of $A_V=0.6$\,mag.
The bottom frame shows a clear difference between a constant SFR
and an instantaneous burst near the 400\,nm break. We can favour
an instantaneous burst scenario based on $K_S$ band data.\label{hyperz}}
\end{figure}

The photometry in optical bands $B_\mathrm{Bessel}$, $V_\mathrm{Bessel}$, 
$R_\mathrm{Cousin}$, $i_\mathrm{Gunn}$, and in infrared bands 
$J_\mathrm{S}$, $H$, and $K_\mathrm{S}$ is described in 
\citet{cabanac05,vallsgabaud06}. Because the deflector is compact in the 
optical range, the source $R_\mathrm{Cousin}$ and $i_\mathrm{Gunn}$ magnitudes were 
simply extracted in an annulus of inner/outer radii of 1\farcsec5/3\arcsec.
The source is faint in $B_\mathrm{Bessel}$ and $V_\mathrm{Bessel}$ bands, 
and we can only put an upper limit to its flux in these bands. The IR magnitudes 
are more difficult to extract because the foreground deflector covers part of 
the source. We subtracted the central lens  with {\sc{galfit}} \citep{peng02},
using the best fit parameters of \citet{cabanac05}. The derived IR magnitudes 
are sensitive to the lens substraction and we estimate that the systematic errors
are of the order of 0.3 mag. We obtain the integrated Spectral Energy Distribution 
(SED, in Vega magnitudes) as shown in Figure~\ref{hyperz}. 

We derived dust extinctions $A_V$, and ages of the stellar populations 
using {\sc{HyperZ}} \citep{bolzonella00}, for the known redshift of 
3.7723, an extinction law similar to the Small Magellanic Cloud 
and several libraries of {\sc{HyperZ}} templates, spanning single burst to 
constant SFR. 
The stellar Initial Mass Function (IMF) is that of \citet{miller79}
and the metallicity of {\sc{HyperZ}} templates is fixed at the solar value.
Although it is likely that FORJ0332-3557 has a subsolar metallicity at 
a look-back time of $\sim12$\,Ga, at face value solar metallicities are as 
good a guess as sub-solar metallicities. In order to estimate the effect of
metallicity on $A_V$, we also fed {\sc{HyperZ}} with the best fit 
{\sc{starburst99}} templates (cf. previous section) of $Z$ = 0.004.

Both Single-burst and Irregular {\sc{HyperZ}} templates yield the same dust 
extinction $A_V=0.5$\,mag (for $R_V=3$, $E[B-V]=0.17$) for the best fit 
23\,Ma-old single-burst template and the best fit 128\,Ma constant SFR 
template (Figure~\ref{hyperz}).  
The ages of the stellar populations are not well constrained by stellar synthesis, 
and Fig.~\ref{hyperz} illustrates this point. 
One template is 100 Ma older than the other and yet they are
 indistinguishable with broadband data only. 
The inferred colour excess of $E(B-V)=0.17$ is similar to the value found by \citet{shapley03}
($0.169\pm0.006$) for the Group 1 (\lya deficient) LGBs, using a similar 
extinction law and photometric bands, but a 300-Ma-old constant star 
formation rate. FORJ0332-3557 also falls in the middle of the extinction distribution 
of the $z\sim4$ LBG sample \citet{steidel99}. 

Following \citet{pettini00}, we infer the extinction at 150\,nm, $A_{150}$ 
and the \Hi\,content, $N($\Hi$)$. For an SMC like extinction law, 
$A_{150}=12.6\times E(B-V) \simeq 2-2.5$. Assuming gas-to-dust 
ratios of$<N($\Hi$)/E(B-V)>=4.93\times 10^{21}$cm$^{-2}$mag$^{-1}$ for the 
Milky Way, $1-2\times 10^{22}$cm$^{-2}$mag$^{-1}$ for LMC,  and
$5\times 10^{22}$cm$^{-2}$mag$^{-1}$ for the SMC, we find 
$7.6\times 10^{20}$cm$^{-2}<N($\Hi$)<10^{21}$cm$^{-2}$ (MW),
$1.5\times 10^{21}$cm$^{-2}<N($\Hi$)<4\times 10^{21}$cm$^{-2}$ (LMC),
and $7.5\times 10^{21}$cm$^{-2}<N($\Hi$)<10^{22}$cm$^{-2}$ (SMC). The
column density inferred from the \lya absorption line is compatible with
the first two values and would be inconsistent with an SMC gas-to-dust ratio.

\subsection{Dust-corrected star formation rate}
The UV (150\,nm) luminosity is a standard indicator of (massive) star formation.
Without correcting for dust extinction and assuming a constant SFR, 
\citet{cabanac05} derived $\mathrm{SFR}_{\mathrm{UV}} 
\approx 31\,(A/12.9)^{-1}$\,\hm2\,\msun a$^{-1}$, where $A$ is 
the gravitational amplification produced by the lens. Using the extinction 
measured in the previous section, $A_{150}=2-2.5$, we get 
$\mathrm{SFR}_{\mathrm{UV}}\approx 196-310\,(A/12.9)^{-1}$\,\hm2\,\msun
a$^{-1}$.  Using the same method on cB58, \citet{pettini00} derives 
$\mathrm{SFR}_{\mathrm{UV}}=395$\,\msun\,a$^{-1}$ (for $H_0=70$\,\kmsmpc,
$q_0=0.1$). This puts FORJ0332-3557 in a very dynamic star forming episode,
similar to what is observed in SDSS J1147$-$0250 \citep{bentz04}. 

\section{Conclusions}
We have presented a medium-resolution spectrum of FORJ0332-3557, 
a lensed $z=3.7723\pm0.0005$ starburst galaxy, magnified 13 times, very 
similar to both $z=2.73$ cB58  and the 
composite spectrum of Lyman-break galaxies at $z \sim 3$.

We find spectral signatures of outflows of $\sim270$\kms, commonly
found in starburst galaxies, through a distinct residual \lya emission 
off-centered by 0\farcsec5 and peaking at ca. $+830$\kms. This
offset emission is most probably produced by the outflow rather than
by an isolated, bright \Hii region within the galaxy.

By combining visible-to-infrared colours and  
spectral features, we derive consistent ages of $\la10$\,Ma for single bursts
and 20-40\,Ma for constant SFRs and an extinction of $A_V=0.5$ using 
two independent stellar populations synthesis codes. Young stellar ages are 
also supported by the detection of the Wolf-Rayet He{\sc ii} $\lambda$164.04\,nm line, 
indicator of stellar ages in the range 4-6\,Ma. There is a clear trend 
to sub-solar metallicities in both cases. The ongoing rate of star formation,
computed from the 150 nm continuum 
is $\mathrm{SFR}_{\mathrm{UV}}\approx 200-300$\,\hm2\,\msun\,a$^{-1}$. 

We also derive preliminary lower limits to the abundances  of some the 
low-ionisation interstellar lines suggesting  sub-solar metallicities in the ionised
gas phase.

\section*{Acknowledgments}
We are grateful to  Max Pettini for providing the spectrum of MS1512-cB58
and important suggestions, to Miguel Cervi\~no for computing a complete set of 
{\tt SED@} templates, to Lise Christensen for useful discussions, and to the
anonymous referee for stimulating comments.

\bibliography{lens2}

\bibliographystyle{mn2e}

\label{lastpage}
\end{document}